\documentclass[entropy,article,accept,moreauthors,pdftex]{Definitions/mdpi}

\graphicspath{{Definitions/}}
\firstpage{1} 
\pubvolume{xx}
\issuenum{1}
\articlenumber{5}
\pubyear{2020}
\copyrightyear{2020}
\history{Received: 24 May 2020; Accepted: 19 June 2020; Published: date}

\usepackage{hhline}
\usepackage{subcaption}
\usepackage{color}

\Title{Automatic Detection of Depression in Speech Using Ensemble Convolutional Neural Networks}

\Author{Adri\'an V\'azquez-Romero 
 and Ascensi\'on Gallardo-Antol\'in\orcidB{}*}

\AuthorNames{Adri\'an V\'azquez-Romero and Ascensi\'on Gallardo-Antol\'in}

\address[1]{%
 Department of Signal Theory and Communications, Universidad Carlos III de Madrid, Avda. de la Universidad, 30, Legan\'es, 28911 Madrid, Spain; avazquez@tsc.uc3m.es}

\corres{Correspondence: gallardo@tsc.uc3m.es}

\abstract{This paper proposes a speech-based method for automatic depression classification. The~system is based on ensemble learning for Convolutional Neural Networks (CNNs) and is evaluated using the data and the experimental protocol provided in the Depression Classification Sub-Challenge (DCC) at the 2016 Audio–Visual Emotion Challenge (AVEC-2016). In the pre-processing phase, speech files are represented as a sequence of log-spectrograms and randomly sampled to balance positive and negative samples. For the classification task itself, first, a more suitable architecture for this task, based on One-Dimensional Convolutional Neural Networks, is built. Secondly, several of these CNN-based models are trained with different initializations and then the corresponding individual predictions are fused by using an Ensemble Averaging algorithm and combined per speaker to get an appropriate final decision. The~proposed ensemble system achieves satisfactory results on the DCC at the AVEC-2016 in comparison with a reference system based on Support Vector Machines and hand-crafted features, with a CNN+LSTM-based system called DepAudionet, and with the case of a single CNN-based classifier.}

\keyword{depression detection; speech;  convolutional neural networks; ensemble learning}

\begin{document}
\section{Introduction}
\label{sec:introduction}
According to the World Health Organization (WHO), over 300 million people are estimated to suffer from depression \cite{world2017depression}, and that number is globally going up, especially at an advanced age. Scientifically known as Major Depressive Disorder (MDD), depression is a mental disorder characterized by a low mood, low self-esteem, loss of interest, low energy, and pain without a clear cause for an extended period of time. It has negative impacts on a person's family, work, sleeping, and eating habits. In extreme cases, as reported by \cite{bachmann2018epidemiology}, half of all completed suicides are related to depressive and other mood disorders. Because of that, many researchers have focused on developing systems to diagnose and prevent this mental disorder to help psychiatrists and psychologists to assist patients as soon as possible.

Traditional mental health disorder reports, such as Beck's Depression Inventory (BDI-II) \cite{beck1988psychometric}, Geriatric Depression Scale (GDS) \cite{yesavage1982development}, Hamilton Rating Scale for Depression (HRSD) \cite{hamilton} and the Patient Health Questionnaire (PHQ-8) \cite{PHQ8} (used as a target variable in this research), are not completely useful or take too much time and consideration for the psychiatrist to diagnose the illness. Moreover, they are quite boring for the patients and could be biased due to that effect.

Automatic Depression Detection (ADD) using computer algorithms is the next step in prevention and monitoring of this mental disorder. Studies on ADD based on speech signals started in the late 2000s~
 \cite{cohn2009detecting} and have taken on increased importance since then. It is worth noting that ADD systems are not intended to substitute medical experts' judgments, but to assist them with clinical decision~support.

In this context, since 2011, annually, Audio–Visual Emotion Challenges have been proposed. The~competitions are \textit{``aimed at comparison of multimedia processing and machine-learning methods for automatic audio, visual, and audio–visual health and emotion sensing, with all participants competing strictly under the same conditions''}. Using different databases, each competition has been focused on one particular mental disorder such as depression \cite{avec2013,avec2014,avec2016} or bipolar disorders \cite{avec2018}, using emotion recognition through four well-known dimensions of the emotions---arousal, valence, expectancy, and power---which have been present in all the challenges \cite{avec2011,avec2013,avec2014,avec2016,avec2018}. Some of the tasks are called \textit{sub-challenges} and researchers can propose solutions for one or many of them.

This research follows the guidelines of the \emph{Depression Classification Sub-Challenge} (DCC) at the 2016 Audio–Visual Emotion Challenge (AVEC) \cite{avec2016}, whose main goal is to determine if a speaker 
has been diagnosed as depressed or not using audio, visual, and/or text data. In particular, in this work, we~focus on audio information. For this task, we propose a system based on ensemble learning whose individual classifiers are Convolutional Neural Networks (CNNs) and the inputs to them are speech log-spectrograms. As in the DCC sub-challenge, the English-speakers database DAIC-WOZ~\cite{daic-woz} is used for the system evaluation.

The~rest of the paper is organized as follows: Section~\ref{sec:related_work} looks over the state of the art of the automatic depression detection problem, CNNs and ensemble methods. In Section~\ref{sec:materials_and_methods}, we present the dataset we have used, and we describe the proposed system. Section~\ref{sec:results_and_discussion} contains the relevant experiments and results. Finally, in Section~\ref{sec:conclusions} we expose our conclusions and some lines of future work.

\section{Related Work}
\label{sec:related_work}
In this section, first, we briefly analyze the systems developed for automatic depression detection in the recent years. After that, an overview of the CNN models and ensemble methods are provided.

Most of the works in the literature related to ADD systems consider two main sources of information, either individual or combined: visual and audio modalities. Although we do not use visual features in this research, we include here a summary of the related state of the art for~completeness.

The~majority of the work on this visual modality is based on facial video information,
by means of the modeling of facial expressions, head movements, eye gazes, or blinks. The~way the information is encoded plays a very important role in these kinds of systems. Facial expressions have been generally represented by Facial Action Coding System (FACS), Active Appearance Model (AAM), or Local Phase Quantization at Three Orthogonal Planes (LPQ-TOP) \cite{jiang2013dynamic}. With these representations, the well-known Support Vector Regression (SVR) models obtain good achievements \cite{cummins2013diagnosis,wen2015automated}. In other works, such as \cite{ooi2011prediction}, eigenfaces are proposed to predict depression in adolescents. Other systems combine visual cues with other types of information, mainly acoustic, resulting in multimodal systems that usually outperform the individual modalities. In this context, it is worth mentioning the system described in \cite{kachele2014fusion} that consists of neural network-based hierarchical classifiers and SVR ensembles for fusing the audio and visual information and was successfully assessed on the AVEC-2013 \cite{avec2013} challenge. Another relevant work is the one described in \cite{yang2016decision} that implements a combination of the audio and visual features with a~Decision Tree as classifier and was the winner of the AVEC-2016 challenge.

As mentioned in Section~\ref{sec:introduction}, in this work we focus on the acoustic modality.
Speech does not only convey linguistic contents, but also contains paralinguistic features (how words are said) that provide important clues about the emotional, neurological, and mental traits and states of the speaker. For~this reason, in recent years, speech technologies are being proposed for the assessment, diagnosis and tracking of different health conditions that affect the subject’s voice \cite{Cummins2018}. In this area, commonly referred to as \textit{Computational Paralinguistic Analysis}, current research encompasses the detection of pathological voices due, for example, to laryngeal disorders \cite{Fang2019}; the diagnosis and monitoring of neurodegenerative conditions, such as Parkinson’s disease \cite{Zlotnik2015, Braga2019}, Mild Cognitive Impairment~\cite{Gosztolya2019}, Alzheimer’s disease~\cite{LopezdeIpina2018, Gosztolya2019} or Amyotrophic Lateral Sclerosis \cite{An2018}; the prediction of stress and cognitive load level \cite{Gallardo-Antolin2019a, Gallardo-Antolin2019b}; and the detection of psychological pathologies, such as autism \cite{Cho2019} or depression~\cite{cummins2015review}, which is the topic of this paper.

Conventional systems for speech-based health tasks consists of data-driven approaches based on hand-crafted acoustic features, such as pitch, prosody, loudness, rate of speech, and energies, among others, and a machine-learning algorithm such as Logistic Regression, Support Vector Machines (SVM) or Gaussian Mixture models \cite{Zlotnik2015, Braga2019, Gosztolya2019, Cho2019}. Nevertheless, very recent works, such as, for example, \cite{Fang2019, LopezdeIpina2018, An2018, Cummins2018, Gallardo-Antolin2019a, Gallardo-Antolin2019b}, deal with the use of deep-learning techniques for these tasks, since, presently, these kinds of methods have achieved unprecedented successes in the field of automatic learning applied to signal processing, and~particularly in image, video, and audio problems.

In the specific field of speech-based automatic depression detection, most of the developed systems also follow one of these two strategies. On the one hand, conventional ADD systems rely on studies about the importance of several acoustic characteristics for depression detection. In fact, according to \cite{cummins2015review,asgari2014inferring,quatieri2012vocal}, the most relevant ones for this task are pitch, formants, intensity, jitter, shimmer, harmonic-to-noise ratio, and rates of speech. These voice quality features are related to the observation that depressed speakers tend to speak in an unnatural and monotonous way \cite{darby1984speech}. On the other hand, recently, several ADD systems based on the deep-learning paradigm have been proposed. In particular, Convolutional Neural Networks (CNN), which represent a specific case of Artificial Neural Network (ANN), are some of the most commonly used architectures in this field.



CNNs appeared in 1980s \cite{fukushima1980neocognitron}, but the research that became a milestone was \cite{krizhevsky2012imagenet} where a CNN-based net was proposed for image recognition tasks. Since then, a large number of problems in computer vision have been solved with this kind of architectures such as handwritten digits, traffic sign classification, people detection, or image recognition for health applications. Later, these~approaches have spread to audio-related tasks, such as for example, automatic speech recognition~\cite{abdel2014convolutional,golik2015convolutional,deng2013recent,Lee2019}, speech emotion recognition \cite{Zheng2019, Hajarolasvadi2019} or acoustic scene classification \cite{piczak2015environmental, Nguyen2018}. For our purposes, the research in \cite{depaudionet} is especially relevant, where a speech-based depression detection system is proposed, called DepAudionet, that uses One-Dimensional CNN (1d-CNN), Long Short-Term Memory (LSTM) and fully connected layers, since it is the basis of our work.  

On the other hand, ensemble methods are meta-algorithms that help to improve machine-learning results by combining more than one model, usually using the same base learner. Ensemble learning with neural networks was introduced in \cite{hansen1990neural} and since then, it has been used in many architectures and applications such as image categorization \cite{kumar2016ensemble}, sentiment analysis \cite{poria2017ensemble} or acoustic environment classification \cite{hwang2016ensemble}.

In this paper, we deal with the problem of automatic depression detection by only using the subject's voice. In particular, we present two main contributions. In the first one, we propose a~refined 1d-CNN architecture based on the aforementioned DepAudionet model \cite{depaudionet} that is optimized by selecting the best configuration from an exhaustive experimentation.
In the second one, we take advantage of the ensemble learning strategy for fusing several machines with this 1d-CNN architecture, in such a way that the performance of these individual systems is improved. Although Ensemble CNN models have been successfully used in other speech- and audio-related tasks, such as automatic speech recognition \cite{Lee2019}, speech emotion recognition \cite{Zheng2019}, or acoustic scene classification \cite{Nguyen2018}, to the best of our knowledge, they have not been previously used for automatic depression detection from~speech. 

An important issue in real-world speech-based health applications is privacy, as patients’ audio and/or video recordings are highly sensitive and contain personal information. In the DAIC-WOZ dataset we have used in this work, this issue does not apply, as participants completed consent forms to allow their data to be shared for research purposes \cite{daic-woz}. However, presently, there is a growth interest in developing this kind of application for real-world scenarios where, at least, the data acquisition is done through microphone-enabled smart devices and/or other wearable technologies that what would allow the remote monitoring of patients \cite{Faurholt-Jepsen2016, Cummins2018}. According to \cite{Low2020}, apart from the consent forms, two main strategies for strengthening privacy can be considered. The~first approach consists of the extraction of acoustic features from which it is not possible to reconstruct the raw speech signal. A~good example is \cite{Little2020}, where the used audio characteristics are the percentage of speech detected in a~given time period, and the percentage of speech uttered by the patient, and therefore, it is not necessary to store the whole raw recordings. When the system uses features that allow the recovery of the original speech signal, a second approach can be applied, consisting of the extraction and encryption of the acoustic characteristics in the local device and their transmission to a secure server where further analysis is done \cite{Faurholt-Jepsen2016}. If the system proposed in this paper had to work in these kinds of real-world scenarios, the second approach should be considered, as its input are log-spectrograms (a reversible transformation of the speech signal), although this issue is beyond the scope of this paper.

\section{Materials and Methods}
\label{sec:materials_and_methods}
\vspace{-6pt}
\subsection{Dataset and Feature Extraction}
\label{sec:dataset}
This section, divided into three subsections, introduces the database used in the research, the~pre-processing done over the audio files and the feature extraction process that converts the speech signals into log-spectrograms with a suitable structure.

\subsubsection{Database Description}
\label{subsec:database_description}
Experiments have been carried out on the Distress Analysis Interview Corpus-Wizard of Oz (DAIC-WOZ) database, a part of the larger corpus DAIC \cite{daic-woz}, which include audio, video, and transcriptions of a questionnaire conducted by an animated virtual interviewer. The~language of the database is English. It is freely available for research purposes at \href{http://dcapswoz.ict.usc.edu}{http://dcapswoz.ict.usc.edu}.

The~DAIC-WOZ dataset contains $189$ subjects distributed in a training set composed of $142$ individuals and a test set consisting of the remaining $47$ ones. 

\begin{figure}[H]
\centering
\includegraphics[width=0.75\textwidth]{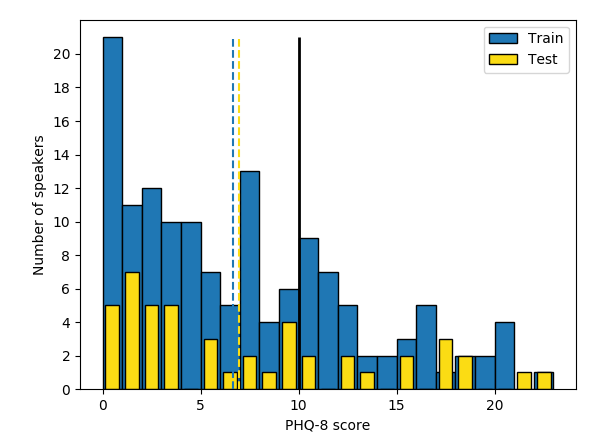}
\caption{Number of speakers per PHQ-8 score for the training and test sets, border between depressed and non-depressed subjects (black continuous line) and the mean of the train and test distributions (discontinuous lines).}
\label{fig:hist_phq8}
\end{figure}

As mentioned before, in this research only the audio files have been taken into account. This~means $189$ audio files ranging between 7--33~min (avg. $16$~minutes). Each patient has one and only one audio in the database and is associated with his/her PHQ-8 score \cite{PHQ8}. PHQ-8 scores in the database are distributed as in the Figure \ref{fig:hist_phq8} where the black vertical line represents the border between the depressed and the non-depressed subjects. This limit has been proposed by the authors of the questionnaire in \cite{PHQ8} and they consider depressed those patients with a greater or equal than 10 score.

Since our objective is classifying between depressed and non-depressed audio recordings, we~transformed the PHQ-8 scores into a binary label (\emph{0} for non-depressed and \emph{1} for depressed) following the \textit{Depression Classification Sub-Challenge} (DCC) \cite{avec2016} guidance. This way, we obtain an imbalanced binary problem with $100$ non-depressed and $42$ depressed audio recordings in the training dataset, one per speaker/patient. The~test set follows a similar distribution, $33$ depressed and $14$~non-depressed~patients.

\subsubsection{Pre-Processing}
\label{subsec:preprocessing}
All audio files contain voices from both the patient and the virtual interviewer, so the first step is to eliminate the second voice because it is computer generated, plain, and emotionless throughout all interviews. In addition, long silences have been removed too. Both tasks have been done by using the diarization algorithm implemented in the Python library \textit{PyAudioAnalysis} \cite{giannakopoulos2015pyaudioanalysis}. Some low-quality recordings were also removed from the training set following the recommendations of the DAIC-WOZ dataset authors. After these processes, there are available $169$ audio files composed by only the speaker voice with a mean length of $547.29$~s and a standard deviation of $237.58$~s. $122$ of these audio recordings are available for training and the remaining $47$ compose the test set.

Due to the small number of audio recordings and their different length, it is a good practice to crop them into shorter segments that we called \emph{samples}, with a same size $S$ in seconds. This technique produces two favorable effects: on the one hand, all the inputs to the classification algorithm have the same size, and, on the other hand, a large number of samples to train the models is generated, and,~therefore these models will be, intuitively, more representative and accurate. As stated in \cite{depaudionet}, $S=4$~s happens to be the best length to crop the files. From now on, we differentiate two~levels: \textit{speaker level} that represents the full audio recording containing voice from a single person who must be classified as depressed or not, and the \textit{sample level} that refers to each of the single crops of $S=4$~s.

The~major problem with the DAIC-WOZ database (and other depression databases) is the uneven sample distribution (see Figure \ref{fig:hist_phq8}) which incurs a large bias in the classification task. In case the model uses this distribution directly to be trained, it will have a big bias to the non-depressed category. Consequently, it is necessary to balance the number of depressed and non-depressed crops of audio but not in a random way, since another bad methodology is using too many samples of a single speaker recording which penalizes the model tendency too. With all of that in mind, the considered sampling technique must maximize the number of $4$~s samples maintaining balance between class labels while selecting the same number of crops for each speaker. 
 
After applying these restrictions, we understand that the optimum number of samples per speaker is $89$~by $62$ speakers in the training set, i.e., $31$ randomly selected speakers from each class, or, what is the same, $5518$ samples of audio to train the model.
In the case of test files, for most of the speakers, we~select $89$ crops, but four speakers have a lower number of crops, since their interventions are shorter than $5$~min and $56$~s. For that reason, the number of samples in the test set is $3991$ preserving samples of all the speakers.

\subsubsection{Feature extraction}
\label{subsec:feature_extraction}
The~DCC at AVEC-2016 \cite{avec2016} provides a baseline with traditional machine-learning techniques (in~particular, a SVM-based classifier) and some hand-crafted speech features extracted with COVAREP (v1.3.2), an open-source MATLAB and Octave toolbox \cite{covarep}: fundamental frequency, formants, energy, normalized amplitude quotient, 24 Mel-Frequency Cepstral Coefficients (MFCC), among others. In this context, other studies \cite{nasir2016multimodal} use the extended Geneva Minimalistic Acoustic Parameter set (eGeMAPS)~\cite{eyben2015geneva}.

However, when dealing with deep-learning approaches in speech-based tasks, a common practice is to consider log-spectrograms or Mel-scale spectrograms  as input features \cite{piczak2015environmental, depaudionet}. In~a~preliminary experimentation, we used these two kinds of acoustic characteristics, yielding similar results, as also reported in \cite{depaudionet} for the depression classification task. Finally, we decided to use the log-spectrogram representation, which is obtained by applying the logarithm to the magnitude of the short-time Fourier transform (STFT) of each input speech crop. The~spectrograms are computed every $32$~ms over Hamming windows of $64$~ms long and setting the number of FFT points to $1024$.

\begin{figure}[H]
\centering
\includegraphics[width=0.4\linewidth]{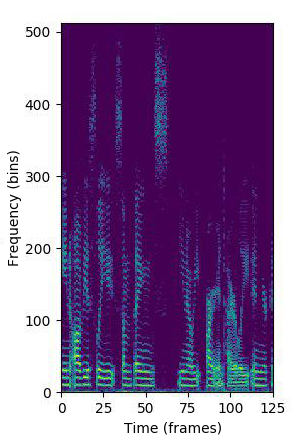}
\caption{Log-spectrogram of a speech crop.}
\label{spectrogram}
\end{figure}

Figure \ref{spectrogram} shows the log-spectrogram of a speech crop, which our model uses as input. Taking into account that the sampling frequency of the audio files is $16$~KHz, the duration of each crop is $S=4$~s and the symmetry of the Fourier transform magnitude for real signals (as is the case of speech), each~log-spectrogram is represented as a matrix with dimensions $F_0$ $\times$ $T_0$, where the frequency $F_0$ and the temporal $T_0$ dimensions are respectively, 513 and 125.

\subsection{Depression Detection System}
\label{sec:model}
The~depression detection system proposed in this paper is composed of two stages: a set of CNN-based classifiers and the fusion of their respective predictions by means of an ensemble method to improve the performance. In this section, first, we describe the architecture of these individual models, and then we detail the ensemble techniques that allow their combination. 

\subsubsection{1d-CNN Model}
\label{subsec:1dCNN_model}
A standard CNN \cite{gurney2014introduction} is a type of Deep Neural Network that includes one or more groups of convolutional and pooling layers. A convolutional layer consists of a set of filters or kernels with small sizes that are repeatedly applied along the full input image. We can interpret the number of filters as the \textit{depth} of the layer.
A pooling layer moves a rectangular window or kernel along the previous layer output and takes a representative number (the maximum, the average, the median, etc.) of the whole window to summarize the information in a smaller frame. This kind of layer produces the output to be smaller than the input and simplifies the next layer training. The~kernels are characterized by three parameters: the \textit{kernel size}, $k$, that measures the window height and width; the stride, $s$, that indicates the step, in pixels or units, the window is moved along the image; and the padding, $p$, that indicates the number of \textit{zeros} we aggregate at the borders of the image to control the spatial size of the output volumes, usually to preserve the spatial size of the input.


To deal with the depression detection problem using CNN models, first we carried out a~set of preliminary experiments with some classical architectures previously used for image analysis, since, in our case, the input to the network is a speech log-spectrogram that, in principle, could be handled as an image. These approaches are considered different squared or rectangular kernels or deeper networks architectures. However, the achieved results with these 2d-CNN architectures were not completely satisfactory.
As pointed in \cite{Deng2013, depaudionet}, a possible explanation for this behavior is that the spatial distribution of the log-spectrogram pixels do not have the same relationships as in an general image representation. In addition, the axes of log-spectrograms do not represent the same magnitude, as correspond to time and frequency dimensions that have very different meanings.

For these reasons, we decided to implement a \textit{One-Dimensional Convolutional Neural Network} (1d-CNN) over the frequency axis instead of using two-dimensional kernels. This architecture is based on the DepAudioNet proposed in \cite{depaudionet} for the same task as ours, but without including the last LSTM layer, because it did not provide significant improvements (see Section \ref{sec:results}) and made the system more complex and, consequently slower.

\begin{figure}[H]
\centering
\includegraphics[width=0.95\textwidth]{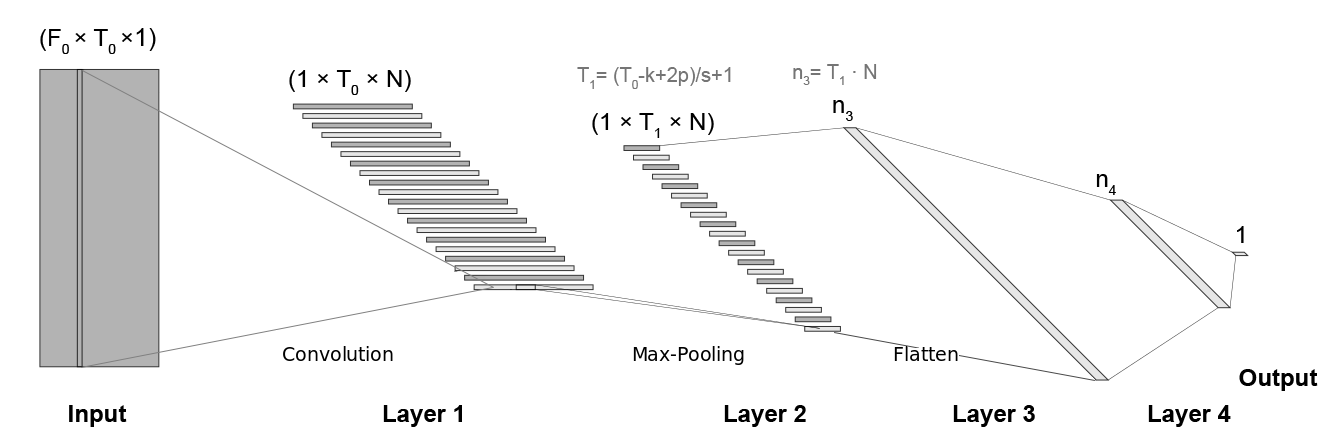}
\caption{CNN architecture. $F_0$ and $T_0$ are the size of the log-spectrogram, $N$ is the number of filters in the 1d-CNN layer, $k$ is the max-pooling kernel size, $s$ is the stride, $p$ is the padding (equal than $s$ in all the experiments), and $n_3$ and $n_4$ are, respectively, the numbers of neurons in the third and fourth~layers.}
\label{fig:cnn_architecture}
\end{figure}

After a preliminary experimentation, finally we adopted the general 1d-CNN architecture shown in Figure \ref{fig:cnn_architecture} that also contains the relationships between the variables that depends on. As can be observed, it is composed of one input layer, four intermediate layers and one output layer.

As mentioned in Section~\ref{subsec:feature_extraction}, the input is a log-spectrogram of $F_0\times T_0$ dimensions. The~first layer is an 1d-CNN whose filters have a $F_0\times 1$ size and cover all the frequency space ($F_0$) and $1$~time slot. We have used different number of filters $N$ in this layer whose results are presented in Section~\ref{sec:results}. The~second layer is a max-pooling layer that moves a window kernel along the time axis and saves the maximum values to represent the whole window, reducing this way its input dimension. The~max-pooling kernel has a size equal to $(1,\; k)$ and it is moved with a stride $s$ of $(1,\; s)$. These two layers use padding to preserve the extremes of the input layer. With this configuration, the model can capture frequency correlations at short-term level through the use of the one-dimensional convolutional layer and temporal dynamics at middle-term level through the use of the one-dimensional pooling layer. 

After that, the max-pooling outputs are flattened and used as input of a fully connected neural network of two layers and one output layer. The~numbers of neurons in these third and fourth layers are $n_3 = T_1 . N$ and $n_4$ respectively, where $T_1$ is the number of temporal slices after the pooling. We~have tried several values for $n_4$ as shown in Section~\ref{sec:results}.

Neurons in all layers have a Rectified Linear Unit (ReLU) as activation function, except the last one, activated by a sigmoid function. The~sigmoid function provides a continuous output in $[0, 1]$ with the probability of each sample to belong to the depressed class.

Please note that at this point, this process is applied at sample level. Let $X_i = \{x_{i,l}\}$, with $l \in \{1, \ldots, L_i\}$ the set of $L_i$ log-spectrograms (samples) corresponding to the $i$th speaker that are the input to the 1d-CNN network. At the output, two sequences are obtained: $P_i = \{p_{i,l}\}$, $l \in \{1, \ldots, L_i\}$, where~$p_{i,l}$ is the probability of the $l$th sample of the $i$th speaker $x_{i,l}$ to belong to the depressed category; and~$Y_i = \{y_{i,l}\}$, $l \in \{1, \ldots, L_i\}$, where $y_{i,l}$ is the binary label assigned to the sample $x_{i,l}$ according to the value of its corresponding probability.

\begin{figure}[H]
\centering
\includegraphics[width=0.95\textwidth]{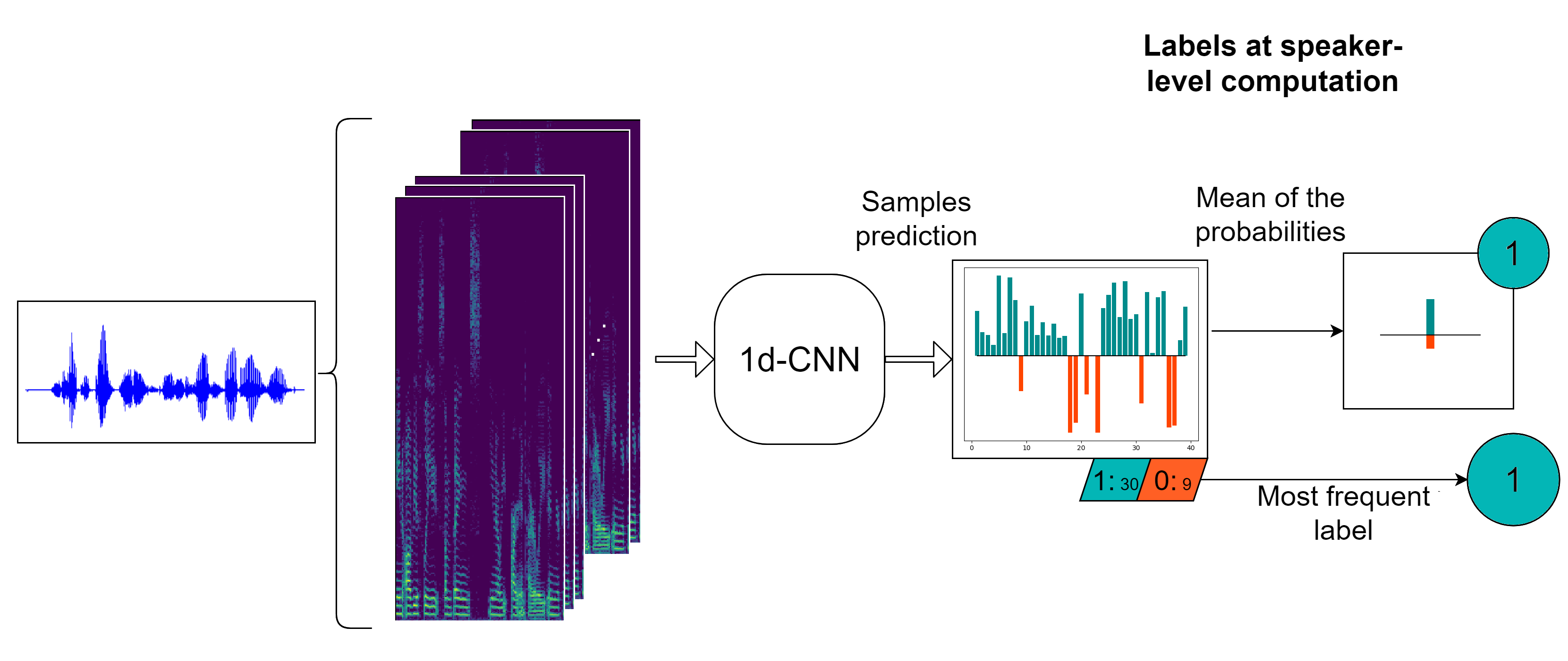}
\caption{Computation of the labels at sample and speaker level.}
\label{fig:predictions}
\end{figure}

As our objective is to classify each speaker as depressed or not, we need to obtain the predictions at speaker level. For doing that, we have considered two different alternatives. The~first one follows the method proposed in \cite{depaudionet}, where the final label assigned to the $i$th speaker is the most frequent prediction value of the samples belonging to this speaker, i.e., the mode of the vector $Y_i$. In this paper, we propose a second mechanism that consists of averaging the vector of sample probabilities $P_i$ corresponding to the $i$th speaker and deciding the final label with a threshold of $0.5$ over this mean probability. Figure \ref{fig:predictions} shows an example of an individual speaker prediction by using both approaches. As can be seen, we extract some log-spectrograms of the raw audio, the network provides a probability and a binary label for each of them and we present the final speaker label with both methods. In most of the cases, they coincide but not always.

\subsubsection{Ensemble Method}
\label{subsec:ensemble_method}
Ensemble methods \cite{berk2006introduction} are meta-algorithms that combine several machine-learning techniques into one predictive model in order to decrease variance (\textit{Bagging}), bias (\textit{Boosting}), or improve predictions~(\textit{Stacking}).

Although these three variants are the most used, in our case, we employ an \textit{Ensemble Averaging} method to combine the sample predictions produced by the different machines (base learners) composing the ensemble. It takes advantage of the randomness in the different base learners, and in consequence, of the differences in the respective predictions.

In our case, all the base learners share the same 1d-CNN architecture described in Section \ref{subsec:1dCNN_model}, and are trained with the same training set but with a different random initialization. Therefore, due~to the difficulties to find the local minimum of the problem, the main goal of the ensemble learning is to take advantage of the different initializations to find local minima and, by means of the fusion of the different 1d-CNN machines outputs, improve both the accuracy and the variance, which is the confidence of the model.

Assuming that $M$ is the number of different 1d-CNN models that compose the ensemble, for~each $i$th speaker, we obtain $M$ probability vectors $P^m_i$, $m \in \{1 \ldots M\}$ and $M$ binary label vectors $Y^m_i$, $m \in \{1 \ldots M\}$ that must be combined in order to obtain the final speaker prediction. To do that, taking into account the two approaches for speaker label computation aforementioned, we propose three different methods of ensemble averaging that are based on the average of the sample probabilities or on the mode of the sample and/or speaker labels.
These three strategies are summarized in a visual representation in Figure \ref{fig:ensemble_architecture} and explained in the following paragraphs: 

\begin{figure}[H]
\centering
\includegraphics[width=0.95\textwidth]{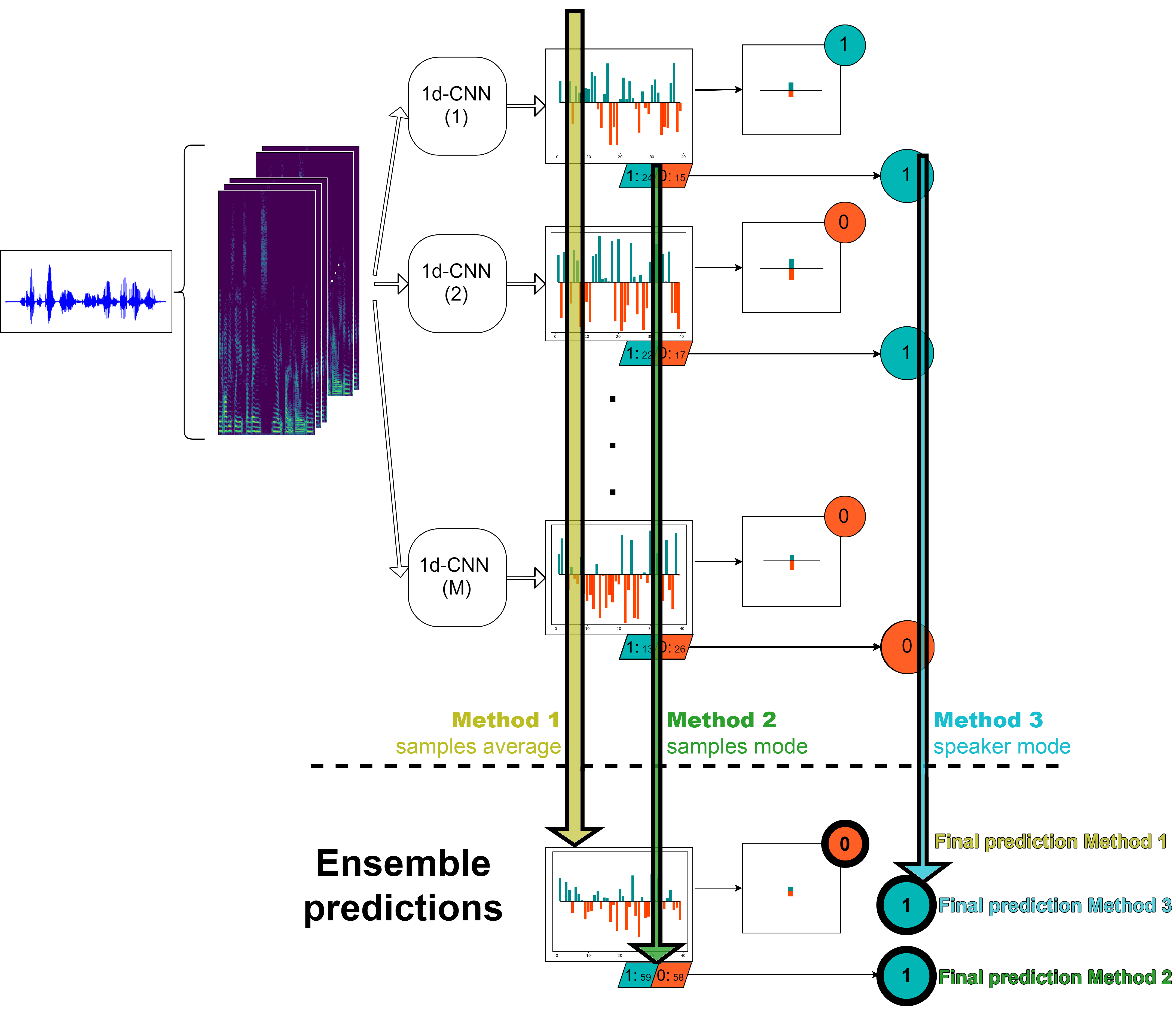}
\caption{Summary of the three proposed ensemble methods.}
\label{fig:ensemble_architecture}
\end{figure}

\begin{itemize}[leftmargin=*,labelsep=5.8mm]
\item \textit{Method 1}. In this alternative, first we average the probabilities at sample level of the $M$ 1d-CNN. After that, as in the case of a single classifier, we compute the probability at speaker level by averaging the probabilities of the samples belonging to this particular speaker and with a~$0.5$~threshold, we are able to decide the final speaker label.

\item \textit{Method 2}. In this strategy, we start from the hard labels at sample level over all the $M$ classifiers and then, we assign to each speaker the most frequent value of all of his/her corresponding $M . L_i$ samples, where $L_i$ is the number of log-spectrograms belonging to the $i$th speaker.

\item \textit{Method 3}. In this approach, first we obtain the predictions at speaker level of each of the $M$ machines as the mode of the corresponding sample labels. Finally, we decide the final speaker label by computing the mode of the $M$ predictions for this particular speaker.
\end{itemize}

In methods where the mode functional is used, in case of $M$ an even number and a tie between the decided labels, a random decision is taken. In addition, note that the average of the probabilities at speaker level could provide a new strategy but it comprises the same commutative operations than~Method~1. 

In Section~\ref{sec:results_and_discussion} a comparison between these three methods is presented. 

\section{Results and Discussion}
\label{sec:results_and_discussion}
\vspace{-6pt}
\subsection{Experimental Protocol}
\label{sec:experimental_protocol}
This section details the tools used to developed the ADD system and the experimental protocol that has been followed to assess it.

To build the different models, the programming language \emph{Python 3.6} has been used. The~computation of log-spectrograms have been possible with the help of the tool \emph{LibROSA} \cite{mcfee2015librosa} and the neural network model is programmed in \emph{Keras} \cite{chollet2015keras}. Others Python libraries used in the experiments and graphs are \emph{Scikit-Learn} \cite{scikit-learn}, \emph{SciPy} \cite{scipy} and \emph{Matplotlib} \cite{Hunter:2007}.

We adopt the F1-score at speaker level as the main metric to assess the system since it is the one proposed in the AVEC-2016 challenge \cite{avec2016}. F1-score is the harmonic average of the precision and recall. The~three of them are more accurate for unbalanced problems than the accuracy that has been calculated too. All of them have been measured both for the depressive and non-depressive speakers.

The~different architectures have been training during $50$ epochs with a batch size of $80$ samples with an \emph{Adadelta} optimizer, a \emph{binary cross-entropy} loss function and a decreasing learning rate from~$1$~to~$0.01$.

Regarding to the system input, each log-spectrogram has been standardized independently using the min-max normalization to obtain log-spectrograms in the range $[0,\; 1]$. Standard normalization has been tried too resulting in a worse performance.

Unless it is stated otherwise, the parameter configuration of the 1d-CNN architecture is as follows: the input dimensions are $F_0 \times T_0 = 513 \times 125$, the number of filters in Layer 1 is $N = 128$ with a $F_0 \times 1$ size, the kernel size, stride and padding in Layer 2 are respectively, $k = 5$, $s = 4$ and $p =4$, and the number of neurons in Layer 4 is $n_4 = 128$.

Concerning the system training, to obtain more reliable results, a 5-fold cross-validation has been used. For each fold, the original training set has been divided into a training and a validation subset, guaranteeing that speakers belong to only one of these subsets. Final metrics have been computed by concatenating the partial results obtained over the test set with the model trained in each one of the 5-fold iterations with the corresponding training+validation configuration.

We compare our proposal to two different reference systems. The~first one is the baseline system provided by the AVEC-2016 challenge that is based on an SVM-based classifier and uses the hand-crafted features mentioned in Section~\ref{subsec:feature_extraction}. The~second one is the DepAudionet depression detection system~\cite{depaudionet}. As mentioned before, its architecture is composed of 1d-CNN, LSTM and fully connected layers with the particular configuration described in \cite{depaudionet}.

\subsection{Results}
\label{sec:results}
In this section, we present the results of the experiments we have carried out to evaluate the performance of the system. 
First, we show the performance of the three ensemble methods under consideration. Next, some parameters of the 1d-CNN architecture have been changed to observe their influence on the performance of the whole system for both the single network and the~ensemble.

\subsubsection{Results with Different Ensemble Methods}
\label{subsec:results_ensemble_methods}
\begin{figure}[H]
\centering
\includegraphics[width=\linewidth]{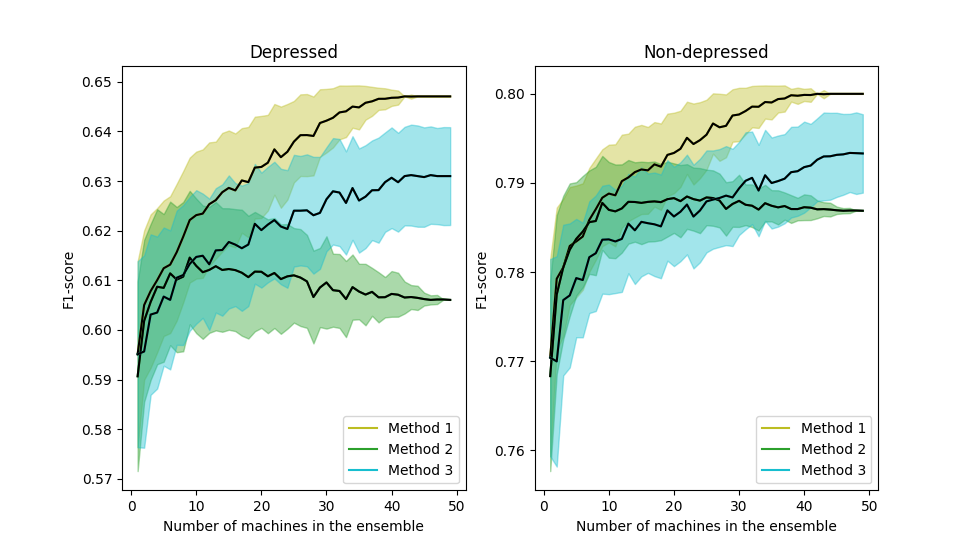}
\caption{F1-score variation as a function of the number of ensemble machines for depressed (\textbf{left}) and non-depressed (\textbf{right}) speakers and the three ensemble methods considered.}
\label{fig:f1_ensemble}
\end{figure}

The~number $M$ of machines that compose the ensemble is quite relevant. Therefore, in this set of experiments we analyze the influence of the value of $M$ on the performance of the system, as well as the improvements achieved by the ensemble averaging algorithm for the three proposed methods: average of the probabilities at sample level (Method 1), most frequent value of the labels at sample level (Method 2) and at speaker level (Method 3).
In particular, Figure \ref{fig:f1_ensemble} represents the variation of the F1-score metric for depressed (left) and non-depressed (right) speakers as a function of the number of ensemble machines, $M$, for the three methods under consideration. The~case of using a single 1d-CNN corresponds to $M=1$. Results for $M > 50$ are not reported, since preliminary experiments showed that values of M greater than $50$ do not improve the performance of the system for this database. The~mean (black line) and the standard deviation (colored shadows) of the F1-score metric have been obtained by taking into account $200$ different combinations of $M=1.50$ machines. 

As can be seen, in general, the performance increases with the number of classifiers in the ensemble for the three methods. Only for the Method 3, there is a worsening in the depressed case when a large number of machines is used. Nevertheless, the results are better than using a single machine. 

In addition, the variance of the individual predictions decreases when the number of machines increases except for the Method 3 (mode of the labels at speaker level). That is due to the big influence of the change of only one speaker prediction in the F1-score metric.
On the contrary, the F1-scores achieved by ensembles performing at sample level are more robust to these variations because the number of samples is considerably higher than the number of speakers.

From these results, we can conclude that the best option is to use the Method 1 and $M = 50$ and then, this is the configuration for the ensemble used in the experiments indicated as \textit{``Ensemble 50 1d-CNN''} in next sections.

\subsubsection{Number of Filters in the Layer 1 and Size of the Layer 4}
\label{results_nofiltersL1_sizeL4}
Table \ref{tab:filters_layer4} presents the four performance metrics of the system varying the number of filters $N$, i.e., the depth of the first layer, and the number of neurons $n_4$ in Layer 4, i.e., the hidden layer of the non-convolutional part. Please note that these two parameters should be related.
In Table \ref{tab:filters_layer4}, it can be seen the results of the AVEC-2016 baseline, the DepAudionet system and both the results of a single 1d-CNN (represented as the mean of $50$ machine performance) and the combination of the same machines with the best ensemble architecture as shown in Section \ref{subsec:results_ensemble_methods}, i.e., Method 1. These metrics are shown for both the depressed and non-depressed classes (in parenthesis).

As expected, the ensemble method provides better F1-scores than the single network. In addition, all the F1-scores obtained by our system in both modalities, single network and ensemble, are better than the baseline. For the depressed class this is because, although the baseline produces a better recall than our system, its precision is clearly worse than those achieved by both the single network and the ensemble. For the non-depressed category, our system also obtains better F1-scores than the baseline, since its recall is clearly higher than the one produced by the baseline, whereas its precision is slightly~lower.
\begin{table}[H]
\centering
\caption{Performance of the system using different number of filters $N$ in Layer 1 and neurons $n_4$ in Layer 4.}
\label{tab:filters_layer4}
\scalebox{0.95}[0.95]{
\begin{tabular}{llllll}
\toprule
$\mathbf{N}$ / $\mathbf{n_4}$ & \textbf{Method}       & \textbf{Accuracy} & \textbf{F1-score} & \textbf{Precision} & \textbf{Recall} \\ \midrule
 &   Baseline (SVM)                    & -  (-)            & 0.41  (0.58)      & 0.27  (0.94) 
 & 0.89  (0.42)    \\ \midrule
 &   DepAudionet                    & 0.65  (0.65)            & 0.50  (0.73)      & 0.44  (0.80) & 0.60  (0.68)    \\ \midrule
\multirow{2}{*}{64 / 64}                                                                    & 1 1d-CNN              & 0.70  (0.70)      & 0.56  (0.77)      & 0.49  (0.83)       & 0.65  (0.71)    \\
                                                                                          & Ensemble 50 1d-CNN & 0.70  (0.70)      & 0.56  (0.77)      & 0.50  (0.83)       & 0.64  (0.73)    \\ \midrule
\multirow{2}{*}{128 / 128}                                                                  & 1 1d-CNN              & 0.70  (0.70)      & 0.59  (0.76)      & 0.50  (0.85)       & 0.72  (0.69)    \\
                                                                                          & Ensemble 50 1d-CNN & 0.74  (0.74)      & 0.65  (0.80)      & 0.55  (0.89)       & 0.79  (0.73)    \\ \midrule
\multirow{2}{*}{256 / 256}                                                                  & 1 1d-CNN              & 0.71  (0.71)      & 0.61  (0.77)      & 0.50  (0.86)       & 0.73  (0.70)    \\
                                                                                          & Ensemble 50 1d-CNN & 0.72  (0.72)      & 0.63  (0.78)      & 0.52  (0.88)       & 0.79  (0.70)    \\ \midrule
\multirow{2}{*}{512 / 512}                                                                  & 1 1d-CNN              & 0.70  (0.70)      & 0.60  (0.76)      & 0.51  (0.86)       & 0.75  (0.68)    \\
                                                                                          & Ensemble 50 1d-CNN & 0.72  (0.72)      & 0.63  (0.78)      & 0.52  (0.88)       & 0.79  (0.70)    \\ \bottomrule
\end{tabular}
}
\end{table}

Regarding the DepAudionet system, again, its performance is better than the baseline in terms of F1-score and precision for the depressed class and in terms of F1-score and recall for the non-depressed class. However, it is outperformed by our systems (single network and ensemble) according to the accuracy, F1-score, precision and recall measures for both the depressed and non-depressed categories.

\begin{figure}[H]
\centering
  \begin{subfigure}[b]{0.48\linewidth}
    \includegraphics[width=\textwidth]{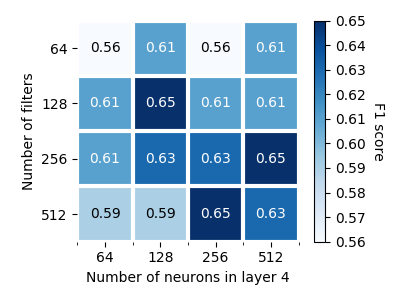}
    \caption{Depressed F1-score}
    \label{fig:1}
  \end{subfigure}
  \begin{subfigure}[b]{0.48\linewidth}
    \includegraphics[width=\textwidth]{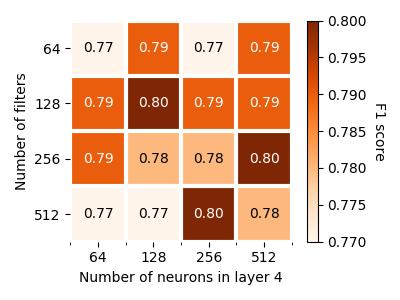}
    \caption{Non-depressed F1-score}
    \label{fig:2}
  \end{subfigure}
  \caption{F1-score variation as a function of the number of filters $N$ in Layer 1 and the number of neurons $n_4$ in Layer 4.}
 \label{grids}
\end{figure}

Figure \ref{grids} shows the F1-scores for more combinations of these $N$ and $n_4$ in a grid representation. It displays the results of an ensemble system with $M=50$ since it is the best one as shown above. Although there are more than one pair of parameters ($N$, $n_4$) that produce the best performance (F1-score of $0.65$ in depressed samples and $0.80$ in non-depressed) we have decided that the best solution is the one with $128$ filters in the convolutional layer and $128$ neurons in the Layer 4, as long as it is the solution with few parameters to train.

\subsection{Size of the Max-Pooling Window}
Another important parameter is the size of the kernel in the max-pooling layer. Taking into account the characteristics of the network, its size is rectangular with a value equal to $1$ in the frequency dimension, due to the one-dimensional filters in the first convolutional layer, and a variable length in the temporal dimension. Consequently, Table \ref{tab:maxpooling} shows a comparative between the kernel lengths at this second layer. Although there are many possibilities changing the stride of the kernels, Table \ref{tab:maxpooling} displays only the best combination for each kernel size.

\begin{table}[H]
\centering
\caption{Performance of the system using different kernels in the max-pooling layer.}
\label{tab:maxpooling}
\scalebox{0.95}[0.95]{
\begin{tabular}{llllll}
\toprule
\textbf{Kernel size}    & \textbf{Method}       & \textbf{Accuracy} & \textbf{F1-score} & \textbf{Precision} & \textbf{Recall} \\ \midrule
          &  Baseline (SVM)                     & -  (-)            & 0.41  (0.58)      & 0.27  (0.94)       & 0.89 (0.42)    \\ \midrule
&   DepAudionet                    & 0.65  (0.65)            & 0.50  (0.73)      & 0.44  (0.80) & 0.60  (0.68)    \\ \midrule
\multirow{2}{*}{(1, 1)} & 1 1d-CNN              & 0.69  (0.69)      & 0.55  (0.76)      & 0.48  (0.82)       & 0.65  (0.71)    \\
                        & Ensemble 50 1d-CNN & 0.68  (0.68)      & 0.55  (0.75)      & 0.47  (0.82)       & 0.64  (0.70)    \\ \midrule
\multirow{2}{*}{(1, 3)} & 1 1d-CNN              & 0.69  (0.69)      & 0.57  (0.75)      & 0.51  (0.84)       & 0.68  (0.72)    \\
                        & Ensemble 50 1d-CNN & 0.72  (0.72)      & 0.61  (0.79)      & 0.53  (0.86)       & 0.71  (0.73)    \\ \midrule
\multirow{2}{*}{(1, 5)} & 1 1d-CNN              & 0.70  (0.70)      & 0.59  (0.76)      & 0.50  (0.85)       & 0.72  (0.69)    \\
                        & Ensemble 50 1d-CNN & 0.74  (0.74)      & 0.65  (0.80)      & 0.55  (0.89)       & 0.79  (0.73)    \\ \midrule
\multirow{2}{*}{(1, 7)} & 1 1d-CNN              & 0.69  (0.69)      & 0.56  (0.75)      & 0.48  (0.84)       & 0.68  (0.69)    \\
                        & Ensemble 50 1d-CNN & 0.72  (0.72)      & 0.61  (0.79)      & 0.53  (0.86)       & 0.71  (0.73)    \\ \bottomrule
\end{tabular}
}
\end{table}

From this table we can see that the best kernel size is the one that takes $5$ temporal slices. For this best case, the stride kernel is ($1$, $4$).

To sum up, the best parameters of the system, at least for this database, are: $N=128$ number of filters, a kernel size in the max-pooling layer of ($1$, $5$) with a stride of ($1$, $4$), using $n_4=128$ neurons at the last hidden layer and $M=50$ number of machines assembled with the Method 1 (i.e., averaging the probabilities at sample level). With these values, our proposed system, the ensemble-based model, achieves a relative improvement in terms of F1-score of $58.5\%$, $30.0\%$ and $10.2\%$ with respect to the baseline, the DepAudionet and the single 1d-CNN architecture, respectively.

\section{Conclusions}
\label{sec:conclusions}
In this paper, we have presented an automatic system for detecting if a person suffers depression by analyzing his/her voice. It is based on an ensemble method, \textit{ensemble averaging}, that combines $M=50$ One-Dimensional Convolutional Neural Networks (1d-CNN).
Each individual 1d-CNN uses log-spectrograms as inputs and is composed of one input layer, four hidden layers and one output layer whose parameters have been optimized by performing an exhaustive experimentation.
Moreover, a random sampling procedure has been used to balance and augment the training data.

The~system has been evaluated over the DAIC-WOZ dataset in the context of the AVEC-2016 Depression Sub-Challenge and compare to the baseline system provided in this challenge, which is based on a SVM-based classifier and hand-crafted features, and the DepAudionet architecture consisting of 1d-CNN, LSTM and fully connected layers. Results have shown that our proposed system achieves a relative improvement in terms of F1-score of $58.5\%$, $30.0\%$ and $10.2\%$ with respect to the baseline, the DepAudionet and the single 1d-CNN architecture, respectively.

For future work, we plan to address the following lines: to study the use of other different methods of ensemble learning as bagging or stacking, to build deeper and narrower networks and the addition of other features as input to the network such as gender or other type of metadata, and to explore different strategies for the combination of our audio-based system with the information provided by the visual modality. 

\vspace{6pt} 

\authorcontributions{conceptualization, A.G.-A.; methodology, A.V.-R. and A.G.-A.; software, A.V.; formal analysis, A.V.-R. and A.G.-A.; investigation, A.V.-R. and A.G.-A.; data curation, A.V.-R.; writing---original draft preparation, A.V.-R.; writing---review and editing, A.G.-A.; supervision, A.G.-A.;funding acquisition, A.G.-A. All~authors have read and agreed to the published version of the manuscript.}

\funding{This research was partly funded by Spanish Government grant TEC2017-84395-P.}


\conflictsofinterest{The~authors declare no conflict of interest.} 

\abbreviations{The~following abbreviations are used in this manuscript:\\

\noindent 
\begin{tabular}{@{}ll}
1d-CNN & One-Dimensional Convolutional Neural Network\\
ADD & Automatic Depression Detection\\
AAM & Active Appearance Model\\
ANN & Artificial Neural Network\\
AVEC & Audio–Visual Emotion Challenge\\
BDI-II & Beck’s Depression Inventory\\
CNN & Convolutional Neural Network\\
DAIC-WOZ & Distress Analysis Interview Corpus - Wizard of Oz\\
DCC & Depression Classification Sub-Challenge\\
eGeMAPS & Geneva Minimalistic Acoustic Parameter Set\\
FACS & Facial Action Coding System\\
GDS & Geriatric Depression Scale\\
HRSD & Hamilton Rating Scale for Depression\\
LPQ-TOP & Local Phase Quantization at Three Orthogonal Planes\\
LSTM & Long Short-Term Memory\\
MDD & Major Depressive Disorder\\
MFCC & Mel-Frequency Cepstral Coefficients\\
PHQ-8 & Patient Health Questionnaire\\
ReLU & Rectified Linear Unit\\
STFT & Short-Time Fourier Transform\\
SVM & Support Vector Machine\\
SVR & Support Vector Regression\\
WHO & World Health Organization
\end{tabular}}

\reftitle{References}



\begin{thebibliography}{999}
\providecommand{\natexlab}[1]{#1}

\bibitem[{World Health Organization} \em{et~al.}(2017){World Health
  Organization} et~al.]{world2017depression}
{World Health Organization}.
\newblock \emph{Depression and Other Common Mental Disorders: Global Health
  Estimates};
\newblock Technical Report; World Health Organization: Geneva, Switzerland, 
 2017.

\bibitem[Bachmann(2018)]{bachmann2018epidemiology}
Bachmann, S.
\newblock Epidemiology of suicide and the psychiatric perspective.
\newblock {\em Int. J. Environ. Res. Public Health} {\bf 2018}, {\em 15},~1425.

\bibitem[Beck \em{et~al.}(1988)Beck, Steer, and Carbin]{beck1988psychometric}
Beck, A.T.; Steer, R.A.; Carbin, M.G.
\newblock Psychometric properties of the {Beck} Depression Inventory:
  Twenty-five years of evaluation.
\newblock {\em Clin. Psychol. Rev.} {\bf 1988}, {\em 8},~77--100.

\bibitem[Yesavage \em{et~al.}(1982)Yesavage, Brink, Rose, Lum, Huang, Adey, and
  Leirer]{yesavage1982development}
Yesavage, J.A.; Brink, T.L.; Rose, T.L.; Lum, O.; Huang, V.; Adey, M.; Leirer,
  V.O.
\newblock Development and validation of a geriatric depression screening scale:
  a preliminary report.
\newblock {\em J. Psychiatr. Res.} {\bf 1982}, {\em 17},~37--49.

\bibitem[Hamilton(1986)]{hamilton}
Hamilton, M.
\newblock The~{Hamilton} rating scale for depression. In {\em Assessment of
  Depression}; Springer: Berlin, Heidelberg, Germany, 
 1986; pp.~143--152.

\bibitem[Kroenke \em{et~al.}(2009)Kroenke, Strine, Spitzer, Williams, Berry,
  and Mokdad]{PHQ8}
Kroenke, K.; Strine, T.W.; Spitzer, R.L.; Williams, J.B.; Berry, J.T.; Mokdad,
  A.H.
\newblock The~{PHQ-8} as a measure of current depression in the general
  population.
\newblock {\em J. Affect. Disord.} {\bf 2009}, {\em 114},~163--173.

\bibitem[Cohn \em{et~al.}(2009)Cohn, Kruez, Matthews, Yang, Nguyen, Padilla,
  Zhou, and De~la Torre]{cohn2009detecting}
Cohn, J.F.; Kruez, T.S.; Matthews, I.; Yang, Y.; Nguyen, M.H.; Padilla, M.T.;
  Zhou, F.; De~la Torre, F.
\newblock Detecting depression from facial actions and vocal prosody.
\newblock  In  Proceedings of the 2009 3rd International Conference on Affective Computing and
  Intelligent Interaction and Workshops, Amsterdam, The Netherlands, 10--12~September~2009; pp. 1--7.

\bibitem[Valstar \em{et~al.}(2013)Valstar, Schuller, Smith, Eyben, Jiang,
  Bilakhia, Schnieder, Cowie, and Pantic]{avec2013}
Valstar, M.; Schuller, B.; Smith, K.; Eyben, F.; Jiang, B.; Bilakhia, S.;
  Schnieder, S.; Cowie, R.; Pantic, M.
\newblock {AVEC}~2013: the continuous audio/visual emotion and depression
  recognition challenge.
\newblock  In~Proceedings of the 3rd ACM International Workshop on Audio/Visual
  Emotion Challenge, Barcelona,~Spain, 21--25~October~2013; ACM: New York, NY, USA, 2013; pp. 3--10.

\bibitem[Valstar \em{et~al.}(2014)Valstar, Schuller, Smith, Almaev, Eyben,
  Krajewski, Cowie, and Pantic]{avec2014}
Valstar, M.; Schuller, B.; Smith, K.; Almaev, T.; Eyben, F.; Krajewski, J.;
  Cowie, R.; Pantic, M.
\newblock {AVEC} 2014: 3d~dimensional affect and depression recognition
  challenge.
\newblock  In Proceedings of the 4th International Workshop on Audio/Visual
  Emotion Challenge, Orlando, FL, USA, 3--7 November 2014; ACM: New~York,~NY,~USA,  2014; pp. 3--10.

\bibitem[Valstar \em{et~al.}(2016)Valstar, Gratch, Schuller, Ringeval, Cowie,
  and Pantic]{avec2016}
Valstar, M.; Gratch, J.; Schuller, B.; Ringeval, F.; Cowie, R.; Pantic, M.
\newblock Summary for {AVEC} 2016: Depression, Mood, and Emotion Recognition
  Workshop and Challenge.
\newblock  In Proceedings of the 24th ACM International Conference on Multimedia, Amsterdam, The Netherlands, 15--19 October 2016;
  ACM: New York, NY, USA,  2016;  pp. 1483--1484.

\bibitem[Ringeval \em{et~al.}(2018)Ringeval, Schuller, Valstar, Cowie, Kaya,
  Schmitt, Amiriparian, Cummins, Lalanne, Michaud, et~al.]{avec2018}
Ringeval, F.; Schuller, B.; Valstar, M.; Cowie, R.; Kaya, H.; Schmitt, M.;
  Amiriparian, S.; Cummins, N.; Lalanne, D.; Michaud, A.; others.
\newblock {AVEC} 2018 workshop and challenge: Bipolar disorder and
  cross-cultural affect recognition.
\newblock  In Proceedings of the 2018 on Audio/Visual Emotion Challenge and
  Workshop, Seoul,~Korea, 22--26 October 2018; ACM: New York, NY, USA,  2018; pp. 3--13.

\bibitem[Schuller \em{et~al.}(2011)Schuller, Valstar, Eyben, McKeown, Cowie,
  and Pantic]{avec2011}
Schuller, B.; Valstar, M.; Eyben, F.; McKeown, G.; Cowie, R.; Pantic, M.
\newblock {AVEC} 2011--the first international audio/visual emotion challenge.
\newblock  In Proceedings of the International Conference on Affective Computing and Intelligent
  Interaction, Memphis,TN, USA, 9--12 October 2011; Springer: Berlin, Heidelberg, Germany, 
 2011; pp. 415--424.

\bibitem[Gratch \em{et~al.}(2014)Gratch, Artstein, Lucas, Stratou, Scherer,
  Nazarian, Wood, Boberg, DeVault, Marsella, Traum, Rizzo, and
  Morency]{daic-woz}
Gratch, J.; Artstein, R.; Lucas, G.; Stratou, G.; Scherer, S.; Nazarian, A.;
  Wood, R.; Boberg, J.; DeVault, D.; Marsella, S.; Traum, D.; Rizzo, A.;
  Morency, L.P.
\newblock The~{Distress} {Analysis} {Interview} {Corpus} of human and computer
  interviews.
\newblock  In Proceedings of the {Ninth} {International} {Conference} on
  {Language} {Resources} and {Evaluation} ({LREC}), Reykjavik, Iceland, 26--31 May  2014; LREC: Reykjavik,
  Iceland,  2014; pp. 3123--3128.

\bibitem[Jiang \em{et~al.}(2013)Jiang, Valstar, Martinez, and
  Pantic]{jiang2013dynamic}
Jiang, B.; Valstar, M.; Martinez, B.; Pantic, M.
\newblock A dynamic appearance descriptor approach to facial actions temporal
  modeling.
\newblock {\em IEEE Trans. Cybern.} {\bf 2013}, {\em
  44},~161--174.

\bibitem[Cummins \em{et~al.}(2013)Cummins, Joshi, Dhall, Sethu, Goecke, and
  Epps]{cummins2013diagnosis}
Cummins, N.; Joshi, J.; Dhall, A.; Sethu, V.; Goecke, R.; Epps, J.
\newblock Diagnosis of depression by behavioural signals: a multimodal
  approach.
\newblock  In Proceedings of the 3rd {ACM} International Workshop on Audio/Visual
  Emotion Challenge, Barcelona, Spain, 21--25 October 2013, pp. 11--20.

\bibitem[Wen \em{et~al.}(2015)Wen, Li, Guo, and Zhu]{wen2015automated}
Wen, L.; Li, X.; Guo, G.; Zhu, Y.
\newblock Automated depression diagnosis based on facial dynamic analysis and
  sparse coding.
\newblock {\em IEEE Trans. Inf. Forensics Secur.} {\bf
  2015}, {\em 10},~1432--1441.

\bibitem[Ooi \em{et~al.}(2011)Ooi, Low, Lech, and Allen]{ooi2011prediction}
Ooi, K.E.B.; Low, L.S.A.; Lech, M.; Allen, N.
\newblock Prediction of clinical depression in adolescents using facial image
  analysis.
\newblock In  Proceedings of the 12th International Workshop on Image Analysis for
  Multimedia Interactive Services ({WIAMIS} 2011), Delft, The Netherlands, 13--15 April 2011; pp. 1--4.

\bibitem[K{\"a}chele \em{et~al.}(2014)K{\"a}chele, Glodek, Zharkov, Meudt, and
  Schwenker]{kachele2014fusion}
K{\"a}chele, M.; Glodek, M.; Zharkov, D.; Meudt, S.; Schwenker, F.
\newblock Fusion of audio-visual features using hierarchical classifier systems
  for the recognition of affective states and the state of depression.
\newblock  In Proceedings of the 3rd International Conference on Pattern
  Recognition Applications and Methods, Angers, France, 6--8~March~2014; pp. 671--678.

\bibitem[Yang \em{et~al.}(2016)Yang, Jiang, He, Pei, Oveneke, and
  Sahli]{yang2016decision}
Yang, L.; Jiang, D.; He, L.; Pei, E.; Oveneke, M.C.; Sahli, H.
\newblock Decision tree based depression classification from audio video and
  language information.
\newblock  In Proceedings of the 6th International Workshop on Audio/Visual
  Emotion Challenge, Amsterdam, The Netherlands,   2016; pp. 89--96.

\bibitem[Cummins \em{et~al.}(2018)Cummins, Baird, and Schuller]{Cummins2018}
Cummins, N.; Baird, A.; Schuller, B.W.
\newblock Speech analysis for health: Current state-of-the-art
  and the increasing impact of deep learning.
\newblock {\em Methods} {\bf 2018}, {\em 151},~41--54,
\newblock
  doi:{\changeurlcolor{black}\href{https://doi.org/https://doi.org/10.1016/j.ymeth.2018.07.007}{\detokenize{10.1016/j.ymeth.2018.07.007}}}.

\bibitem[Fang \em{et~al.}(2019)Fang, Tsao, Hsiao, Chen, Lai, Lin, and
  Wang]{Fang2019}
Fang, S.H.; Tsao, Y.; Hsiao, M.J.; Chen, J.Y.; Lai, Y.H.; Lin, F.C.; Wang, C.T.
\newblock Detection of pathological voice using cepstrum
  vectors: a deep learning approach.
\newblock {\em J. Voice} {\bf 2019}, {\em 33},~634--641,
\newblock
  doi:{\changeurlcolor{black}\href{https://doi.org/https://doi.org/10.1016/j.jvoice.2018.02.003}{\detokenize{10.1016/j.jvoice.2018.02.003}}}.

\bibitem[Zlotnik \em{et~al.}(2015)Zlotnik, Montero, San-Segundo, and
  Gallardo-Antol\'in]{Zlotnik2015}
Zlotnik, A.; Montero, J.M.; San-Segundo, R.; Gallardo-Antol\'in, A.
\newblock Random Forest-based prediction of Parkinson's
  disease progression using acoustic, ASR and intelligibility features.
\newblock  \emph{Proc. Interspeech} \textbf{2015},  \emph{2015}, 503--507.

\bibitem[Braga \em{et~al.}(2019)Braga, Madureira, Coelho, and Ajith]{Braga2019}
Braga, D.; Madureira, A.M.; Coelho, L.; Ajith, R.
\newblock Automatic detection of Parkinson’s disease based
  on acoustic analysis of speech.
\newblock {\em Eng. Appl. Artif. Intell.} {\bf 2019},
  {\em 77},~148--158,
\newblock
  doi:{\changeurlcolor{black}\href{https://doi.org/https://doi.org/10.1016/j.engappai.2018.09.018}{\detokenize{10.1016/j.engappai.2018.09.018}}}.

\bibitem[Gosztolya \em{et~al.}(2019)Gosztolya, Vincze, T\'oth, P\'ak\'aski,
  K\'alm\'an, and Hoffmann]{Gosztolya2019}
Gosztolya, G.; Vincze, V.; T\'oth, L.; P\'ak\'aski, M.; K\'alm\'an, J.;
  Hoffmann, I.
\newblock Identifying Mild Cognitive Impairment and mild
  Alzheimer’s disease based on spontaneous speech using ASR and linguistic
  features.
\newblock {\em Comput. Speech Lang.} {\bf 2019}, {\em 53},~181--197,
\newblock
  doi:{\changeurlcolor{black}\href{https://doi.org/https://doi.org/10.1016/j.csl.2018.07.007}{\detokenize{10.1016/j.csl.2018.07.007}}}.

\bibitem[Lopez-de Ipina \em{et~al.}(2018)Lopez-de Ipina, Martinez-de Lizarduy,
  Calvo, Mekyska, Beitia, Barroso, Estanga, Tainta, and
  Ecay-Torres]{LopezdeIpina2018}
Lopez-de Ipina, K.; Martinez-de Lizarduy, U.; Calvo, P.M.; Mekyska, J.; Beitia,
  B.; Barroso, N.; Estanga, A.; Tainta, M.; Ecay-Torres, M.
\newblock Advances on automatic speech analysis for early
  detection of Alzheimer disease: A non-linear multi-task approach.
\newblock {\em Curr. Alzheimer Res.} {\bf 2018}, {\em 15},~139--148,
\newblock
  doi:{\changeurlcolor{black}\href{https://doi.org/https://doi.org/10.2174/1567205014666171120143800}{\detokenize{10.2174/1567205014666171120143800}}}.

\bibitem[An \em{et~al.}(2018)An, Kim, Teplansky, Green, Campbell, Yunusova,
  Heitzman, and Wang]{An2018}
An, K.; Kim, M.; Teplansky, K.; Green, J.; Campbell, T.; Yunusova, Y.;
  Heitzman, D.; Wang, J.
\newblock Automatic early detection of amyotrophic lateral
  sclerosis from intelligible speech using convolutional neural networks.
\newblock  \emph{Proc.~Interspeech} \textbf{2018}, \emph{2018}, 1913--1917,
\newblock
  doi:{\changeurlcolor{black}\href{https://doi.org/10.21437/Interspeech.2018-2496}{\detokenize{10.21437/Interspeech.2018-2496}}}.

\bibitem[{Gallardo-Antol\'in} and
  {Montero}(2019{\natexlab{a}})]{Gallardo-Antolin2019a}
{Gallardo-Antol\'in}, A.; {Montero}, J.M.
\newblock A saliency-based attention {LSTM} model for
  cognitive load classification from speech.
\newblock  \emph{Proc. Interspeech} \textbf{2019},  \emph{2019}, 216--220,
\newblock
  doi:{\changeurlcolor{black}\href{https://doi.org/10.21437/Interspeech.2019-1603}{\detokenize{10.21437/Interspeech.2019-1603}}}.

\bibitem[{Gallardo-Antol\'in} and
  {Montero}(2019{\natexlab{b}})]{Gallardo-Antolin2019b}
{Gallardo-Antol\'in}, A.; {Montero}, J.M.
\newblock External attention {LSTM} models for cognitive load
  classification from speech.
\newblock {\em Lect. Notes Comput. Sci. (Incl. Subser. Lect. Notes Artif. Intell. Lect. Notes Bioinform.)} {\bf
  2019},~{\em 11816},~139--150.
\newblock
  doi:{\changeurlcolor{black}\href{https://doi.org/10.1007/978-3-030-31372-2_12}{\detokenize{10.1007/978-3-030-31372-2_12}}}.

\bibitem[Cho \em{et~al.}(2019)Cho, Liberman, Ryant, Cola, Schultz, and
  Parish-Morris]{Cho2019}
Cho, S.; Liberman, M.; Ryant, N.; Cola, M.; Schultz, R.T.; Parish-Morris, J.
\newblock Automatic detection of autism spectrum disorder in
  children using acoustic and text features from brief natural conversations.
\newblock  \emph{Proc.~Interspeech} \textbf{2019},  \emph{2019}, 2513--2517,
\newblock
  doi:{\changeurlcolor{black}\href{https://doi.org/10.21437/Interspeech.2019-1452}{\detokenize{10.21437/Interspeech.2019-1452}}}.

\bibitem[Cummins \em{et~al.}(2015)Cummins, Scherer, Krajewski, Schnieder, Epps,
  and Quatieri]{cummins2015review}
Cummins, N.; Scherer, S.; Krajewski, J.; Schnieder, S.; Epps, J.; Quatieri,
  T.F.
\newblock A review of depression and suicide risk assessment using speech
  analysis.
\newblock {\em Speech Commun.} {\bf 2015}, {\em 71},~10--49.

\bibitem[Asgari \em{et~al.}(2014)Asgari, Shafran, and
  Sheeber]{asgari2014inferring}
Asgari, M.; Shafran, I.; Sheeber, L.B.
\newblock Inferring clinical depression from speech and spoken utterances.
\newblock  In~Proceedings of the 2014 {IEEE} International Workshop on Machine Learning for Signal
  Processing (MLSP), Reims, France,  21--24 September 2014; pp. 1--5.

\bibitem[Quatieri and Malyska(2012)]{quatieri2012vocal}
Quatieri, T.F.; Malyska, N.
\newblock Vocal-source biomarkers for depression: A link to psychomotor
  activity.
\newblock In~Proceedings of the  Thirteenth Annual Conference of the International Speech
  Communication Association, Portland, Oregon, 9--13 September 2012.

\bibitem[Darby \em{et~al.}(1984)Darby, Simmons, and Berger]{darby1984speech}
Darby, J.K.; Simmons, N.; Berger, P.A.
\newblock Speech and voice parameters of depression: A pilot study.
\newblock {\em J.~Commun.~Disord.} {\bf 1984}, {\em
  17},~75--85.

\bibitem[Fukushima(1980)]{fukushima1980neocognitron}
Fukushima, K.
\newblock Neocognitron: A self-organizing neural network model for a mechanism
  of pattern recognition unaffected by shift in position.
\newblock {\em Biol. Cybern.} {\bf 1980}, {\em 36},~193--202.

\bibitem[Krizhevsky \em{et~al.}(2012)Krizhevsky, Sutskever, and
  Hinton]{krizhevsky2012imagenet}
Krizhevsky, A.; Sutskever, I.; Hinton, G.E.
\newblock Imagenet classification with deep convolutional neural networks.
\newblock  In \emph{Advances in Neural Information Processing Systems}; Curran Associates Inc.: Red Hook, NY, USA, 2012; 
 pp.
  1097--1105.

\bibitem[Abdel-Hamid \em{et~al.}(2014)Abdel-Hamid, Mohamed, Jiang, Deng, Penn,
  and Yu]{abdel2014convolutional}
Abdel-Hamid, O.; Mohamed, A.r.; Jiang, H.; Deng, L.; Penn, G.; Yu, D.
\newblock Convolutional neural networks for speech recognition.
\newblock {\em IEEE/ACM Trans. Audio Speech Lang. Process.}
  {\bf 2014}, {\em 22},~1533--1545.

\bibitem[Golik \em{et~al.}(2015)Golik, T{\"u}ske, Schl{\"u}ter, and
  Ney]{golik2015convolutional}
Golik, P.; T{\"u}ske, Z.; Schl{\"u}ter, R.; Ney, H.
\newblock Convolutional neural networks for acoustic modeling of raw time
  signal in {LVCSR}.
\newblock In  Proceedings of the  Sixteenth Annual Conference of the International Speech
  Communication Association, Dresden, Germany, 6--10 September 2015.

\bibitem[Deng \em{et~al.}(2013)Deng, Li, Huang, Yao, Yu, Seide, Seltzer, Zweig,
  He, Williams, et~al.]{deng2013recent}
Deng, L.; Li, J.; Huang, J.T.; Yao, K.; Yu, D.; Seide, F.; Seltzer, M.; Zweig,
  G.; He, X.; Williams, J.; others.
\newblock Recent advances in deep learning for speech research at {Microsoft}.
\newblock In  Proceedings of the  2013 {IEEE} International Conference on Acoustics, Speech, and
  Signal Processing ({ICASSP} 2013), Vancouver,~BC,~Canada,  26--31~May~2013; pp. 8604--8608.

\bibitem[Lee \em{et~al.}(2019)Lee, Lee, and Chang]{Lee2019}
Lee, M.; Lee, J.; Chang, J.H.
\newblock Ensemble of jointly trained deep neural network-based acoustic models
  for reverberant speech recognition.
\newblock {\em Digit. Signal Process.} {\bf 2019}, {\em 85},~1--9.
\newblock
  doi:{\changeurlcolor{black}\href{https://doi.org/https://doi.org/10.1016/j.dsp.2018.11.005}{\detokenize{10.1016/j.dsp.2018.11.005}}}.

\bibitem[Zheng \em{et~al.}(2019)Zheng, Wang, and Jia]{Zheng2019}
Zheng, C.; Wang, C.; Jia, N.
\newblock An ensemble model for multi-level speech emotion recognition.
\newblock {\em Appl. Sci.} {\bf 2019}, {\em 10},~205,
\newblock
  doi:{\changeurlcolor{black}\href{https://doi.org/10.3390/app10010205}{\detokenize{10.3390/app10010205}}}.

\bibitem[Hajarolasvadi and Demirel(2019)]{Hajarolasvadi2019}
Hajarolasvadi, N.; Demirel, H.
\newblock 3{D} {CNN}-based speech emotion recognition using k-means clustering
  and spectrograms.
\newblock {\em Entropy} {\bf 2019}, {\em 21}, 479,
\newblock
  doi:{\changeurlcolor{black}\href{https://doi.org/10.3390/e21050479}{\detokenize{10.3390/e21050479}}}.

\bibitem[Piczak(2015)]{piczak2015environmental}
Piczak, K.J.
\newblock Environmental sound classification with convolutional neural
  networks.
\newblock  In Proceedings of the 2015 IEEE 25th International Workshop on Machine Learning for Signal
  Processing (MLSP), Boston,~MA,~USA, 17--20 September 2015; pp. 1--6.

\bibitem[Nguyen and Pernkopf(2018)]{Nguyen2018}
Nguyen, T.; Pernkopf, F.
\newblock Acoustic scene classification using a convolutional neural network
  ensemble and nearest neighbor filters.
\newblock  In Proceedings of the Detection and Classification of Acoustic Scenes
  and Events 2018 Workshop ({DCASE2}018), Surrey, UK, 19--20 November 2018; pp. 34--38.

\bibitem[Ma \em{et~al.}(2016)Ma, Yang, Chen, Huang, and Wang]{depaudionet}
Ma, X.; Yang, H.; Chen, Q.; Huang, D.; Wang, Y.
\newblock DepAudioNet: An Efficient Deep Model for Audio Based Depression
  Classification.
\newblock  Proceedings of the 6th International Workshop on Audio/Visual
  Emotion Challenge;  Amsterdam, The Netherlands, 16 October~2016; 
 ACM: New York, NY, USA, 2016; {AVEC} '16, pp. 35--42.

\bibitem[Hansen and Salamon(1990)]{hansen1990neural}
Hansen, L.K.; Salamon, P.
\newblock Neural network ensembles.
\newblock {\em IEEE Trans. Pattern Anal. Mach. Intell.}
  {\bf 1990},~{\em 12}, 993--1001.

\bibitem[Kumar \em{et~al.}(2016)Kumar, Kim, Lyndon, Fulham, and
  Feng]{kumar2016ensemble}
Kumar, A.; Kim, J.; Lyndon, D.; Fulham, M.; Feng, D.
\newblock An ensemble of fine-tuned convolutional neural networks for medical
  image classification.
\newblock {\em IEEE J. Biomed. Health Inform.} {\bf 2016},
  {\em 21},~31--40.

\bibitem[Poria \em{et~al.}(2017)Poria, Peng, Hussain, Howard, and
  Cambria]{poria2017ensemble}
Poria, S.; Peng, H.; Hussain, A.; Howard, N.; Cambria, E.
\newblock Ensemble application of convolutional neural networks and multiple
  kernel learning for multimodal sentiment analysis.
\newblock {\em Neurocomputing} {\bf 2017},~{\em 261}, 217--230.

\bibitem[Hwang \em{et~al.}(2016)Hwang, Park, and Chang]{hwang2016ensemble}
Hwang, I.; Park, H.M.; Chang, J.H.
\newblock Ensemble of deep neural networks using acoustic environment
  classification for statistical model-based voice activity detection.
\newblock {\em Comput. Speech Lang.} {\bf 2016}, {\em 38},~1--12.

\bibitem[Faurholt-Jepsen \em{et~al.}(2016)Faurholt-Jepsen, Busk, Frost,
  Vinberg, Christensen, Winther, Bardram, and Kessing]{Faurholt-Jepsen2016}
Faurholt-Jepsen, M.; Busk, J.; Frost, M.; Vinberg, M.; Christensen, E.M.;
  Winther, O.; Bardram, J.E.; Kessing, L.V.
\newblock Voice analysis as an objective state marker in
  bipolar disorder.
\newblock {\em Transl. Psychiatry} {\bf 2016}, {\em 6}, e856.
\newblock
  doi:{\changeurlcolor{black}\href{https://doi.org/10.1038/tp.2016.123}{\detokenize{10.1038/tp.2016.123}}}.

\bibitem[Low \em{et~al.}(2020)Low, Bentley, and Ghosh]{Low2020}
Low, D.M.; Bentley, K.H.; Ghosh, S.S.
\newblock Automated assessment of psychiatric disorders using
  speech: A~systematic review.
\newblock {\em Laryngoscope Investig. Otolaryngol.} {\bf 2020}, {\em
  5},~96--116,
\newblock
  doi:{\changeurlcolor{black}\href{https://doi.org/10.1002/lio2.354}{\detokenize{10.1002/lio2.354}}}.

\bibitem[Little \em{et~al.}(2020)Little, Alshabrawy, Stow, Ferrier, McNaney,
  Jackson, Ladha, Ladha, Ploetz, Bacardit, Olivier, Gallagher, and
  O'Brien]{Little2020}
Little, B.; Alshabrawy, O.; Stow, D.; Ferrier, I.N.; McNaney, R.; Jackson,
  D.G.; Ladha, K.; Ladha,~C.; Ploetz, T.; Bacardit, J.; Olivier, P.; Gallagher,
  P.; O'Brien, J.T.
\newblock Deep learning-based automated speech detection as a
  marker of social functioning in late-life depression.
\newblock {\em Psychol. Med.} {\bf 2020}, 1--10.
\newblock
  doi:{\changeurlcolor{black}\href{https://doi.org/10.1017/S0033291719003994}{\detokenize{10.1017/S0033291719003994}}}.

\bibitem[Giannakopoulos(2015)]{giannakopoulos2015pyaudioanalysis}
Giannakopoulos, T.
\newblock pyAudioAnalysis: An Open-Source {Python} Library for Audio Signal
  Analysis.
\newblock {\em PLoS~ONE} {\bf 2015}, {\em 10}, 0144610.

\bibitem[Degottex \em{et~al.}(2014)Degottex, Kane, Drugman, Raitio, and
  Scherer]{covarep}
Degottex, G.; Kane, J.; Drugman, T.; Raitio, T.; Scherer, S.
\newblock {COVAREP}, A collaborative voice analysis repository for speech
  technologies.
\newblock  In Proceedings of the 2014 {IEEE} international conference on acoustics, speech and signal
  processing (ICASSP), Florence, Italy, 4--9 May 2014; pp. 960--964.

\bibitem[Nasir \em{et~al.}(2016)Nasir, Jati, Shivakumar, Nallan~Chakravarthula,
  and Georgiou]{nasir2016multimodal}
Nasir, M.; Jati, A.; Shivakumar, P.G.; Nallan~Chakravarthula, S.; Georgiou, P.
\newblock Multimodal and multiresolution depression detection from speech and
  facial landmark features.
\newblock  In Proceedings of the 6th International Workshop on Audio/Visual
  Emotion Challenge; Amsterdam, The Netherlands, 16 October~2016; 
 pp. 43--50.

\bibitem[Eyben \em{et~al.}(2015)Eyben, Scherer, Schuller, Sundberg, Andr{\'e},
  Busso, Devillers, Epps, Laukka, Narayanan, et~al.]{eyben2015geneva}
Eyben, F.; Scherer, K.R.; Schuller, B.W.; Sundberg, J.; Andr{\'e}, E.; Busso,
  C.; Devillers, L.Y.; Epps, J.; Laukka, P.; Narayanan, S.S.; et al.
\newblock The~{Geneva} minimalistic acoustic parameter set ({GeMAPS}) for voice
  research and affective computing.
\newblock {\em IEEE Trans. Affect. Comput.} {\bf 2015}, {\em
  7},~190--202.

\bibitem[Gurney(2014)]{gurney2014introduction}
Gurney, K.
\newblock {\em An Introduction to Neural Networks}; CRC Press: London, UK, 
  2014.

\bibitem[Deng \em{et~al.}(2013)Deng, Abdel-Hamid, and Yu]{Deng2013}
Deng, L.; Abdel-Hamid, O.; Yu, D.
\newblock A deep convolutional neural network using heterogeneous pooling for
  trading acoustic invariance with phonetic confusion.
\newblock  In Proceedings of the 2013 {IEEE} International Conference on Acoustics, Speech, and
  Signal Processing ({ICASSP} 2013),  Vancouver,~BC,~Canada, 26--31~May~2013; pp. 6669--6673.
\newblock
  doi:{\changeurlcolor{black}\href{https://doi.org/10.1109/ICASSP.2013.6638952}{\detokenize{10.1109/ICASSP.2013.6638952}}}.

\bibitem[Berk(2006)]{berk2006introduction}
Berk, R.A.
\newblock An introduction to ensemble methods for data analysis.
\newblock {\em Sociol. Methods Res.} {\bf 2006}, {\em
  34},~263--295.

\bibitem[McFee \em{et~al.}(2015)McFee, Raffel, Liang, Ellis, McVicar,
  Battenberg, and Nieto]{mcfee2015librosa}
McFee, B.; Raffel, C.; Liang, D.; Ellis, D.P.; McVicar, M.; Battenberg, E.;
  Nieto, O.
\newblock {LibROSA}: Audio and music signal analysis in {Python}.
\newblock  In Proceedings of the 14th {Python} in Science Conference, Austin, Texas, 6--12~July~2015; pp.
  18--25.

\bibitem[Chollet \em{et~al.}(2015)Chollet et~al.]{chollet2015keras}
Chollet, F.;
\newblock Keras. 2015.
\newblock Available online: \url{https://keras.io} (accessed on 19 June 2020). 
 

\bibitem[Pedregosa \em{et~al.}(2011)Pedregosa, Varoquaux, Gramfort, Michel,
  Thirion, Grisel, Blondel, Prettenhofer, Weiss, Dubourg, Vanderplas, Passos,
  Cournapeau, Brucher, Perrot, and Duchesnay]{scikit-learn}
Pedregosa, F.; Varoquaux, G.; Gramfort, A.; Michel, V.; Thirion, B.; Grisel,
  O.; Blondel, M.; Prettenhofer,~P.; Weiss, R.; Dubourg, V.; et al.
\newblock Scikit-learn: Machine Learning in {P}ython.
\newblock {\em J. Mach. Learn. Res.} {\bf 2011},~{\em
  12},~2825--2830.

\bibitem[Jones \em{et~al.}(2001)Jones, Oliphant, Peterson, et~al.]{scipy}
Jones, E.; Oliphant, T.; Peterson, P.;
\newblock {SciPy}: Open Source Scientific Tools for {Python}. 2001.
\newblock Available online: \url{https://www.scipy.org} (accessed on 19 June 2020). 


\bibitem[Hunter(2007)]{Hunter:2007}
Hunter, J.D.
\newblock Matplotlib: A {2D} graphics environment.
\newblock {\em Comput. Sci. Eng.} {\bf 2007}, {\em
  9},~90--95.
\newblock
  doi:{\changeurlcolor{black}\href{https://doi.org/10.1109/MCSE.2007.55}{\detokenize{10.1109/MCSE.2007.55}}}.

\end{thebibliography}

\end{document}